# Experimental verification of magnetic near-field channeling using a helical-shaped wire medium lens

Tiago A. Morgado[1,*], Guilherme L. João[1], Ricardo A. M. Pereira[2], David E. Fernandes[3] and Sylvain Lannebère[1]

[1]*Instituto de Telecomunicações and Department of Electrical Engineering, University of Coimbra, 3030-290 Coimbra, Portugal*

[2]*Instituto de Telecomunicações and Department of Electronics, Telecommunications and Informatics, University of Aveiro, 3810-193 Aveiro, Portugal*

[3]*Instituto de Telecomunicações, Avenida Rovisco Pais, 1, 1049-001 Lisboa, Portugal*

E-mail: tiago.morgado@co.it.pt

**Abstract**

We experimentally verify that a magnetic uniaxial wire medium lens consisting of a racemic array of helical-shaped metallic wires may enable channeling the normal component of the magnetic field of near-field sources with resolution well below the diffraction limit over a broad bandwidth. It is experimentally demonstrated that the helical-shaped wire medium lens can be regarded as the magnetic counterpart of the usual wire medium lenses formed by straight metallic wires. The experimental results are validated with full-wave numerical simulations. We envision that the proposed metamaterial lens may have potential applications in magnetic resonance imaging (MRI), near-field wireless power transfer (NF-WPT), and sensing.

---

[*] To whom correspondence should be addressed: E-mail: tiago.morgado@co.it.pt

The spatial resolution of conventional imaging devices is limited to about one half-wavelength, known as the diffraction limit. Manipulating the near field to overcome the diffraction limit is one of the most exciting applications of metamaterials. Various metamaterial-based subwavelength imaging systems have been proposed and discussed over the last two decades [1-15]. A particularly interesting mechanism is the one based on uniaxial arrays of parallel metallic wires that operate as almost perfect endoscopes and enable a pixel-by-pixel subwavelength channeling of the entire source-radiation spatial spectrum (the so-called "canalization regime") [7-9,16-18]. Yet, such wire medium lenses have an important polarization constraint as they only enable near-field imaging of transverse magnetic (TM) or *p*-polarized waves (magnetic field is parallel to the interface). For transverse electric (TE) or *s*-polarized waves (electric field is parallel to the interface), the wire medium lenses are fully transparent.

A solution to overcome the polarization sensitivity of the uniaxial wire medium lenses and fully restore the electric field radiated by a near-field source with arbitrary polarization was theoretically suggested [19] and experimentally demonstrated [20] some time ago. The proposed approach is based on the post-processing of three linearly independent measurements of the electric field using a metamaterial lens formed by an array of tilted metallic wires. However, such a solution only works for sources that radiate time-stationary fields.

A different possibility to achieve a channeling effect for TE polarized waves in wire medium lenses is by replacing the metallic wires with "perfectly magnetic conducting" (PMC) wires. Nonetheless, broadband PMC materials are mostly a theoretical idealization since they are not readily available in Nature. Fortunately, it may be possible to exploit the unusual electromagnetic responses of artificial microstructured materials to engineer "magnetic wires". In Ref. [10] it was shown that arrays of "Swiss rolls" – periodic arrangements of cylinders comprising a thin conducting sheet wrapped around a central mandrel – provide a viable path to create PMC wires at MHz frequencies.

Sometime ago, we proposed an alternative way to imitate the response of PMC wires that can be implemented in the microwave regime and even possibly extended to higher frequency bands, based on a metamaterial consisting of a racemic array of helical-shaped metallic wires (designated as "magnetic uniaxial wire medium" [21]) [see Fig. 1(a)-(b)]. We theoretically and numerically demonstrated in Ref. [21] that such a metamaterial may be regarded to some extent as the magnetic analogue of the standard wire medium formed by straight metallic wires. In particular, it was shown that under a TE excitation, the magnetic uniaxial wire medium supports a quasi-transverse electromagnetic (q-TEM) mode with phase velocity nearly independent of the transverse wave vector, similar to the mode supported by the standard wire medium for TM incident waves [7-9]. Moreover, we proved numerically that the magnetic uniaxial wire medium enables channeling the near field of TE polarized waves. The objective of this work is to experimentally confirm the findings of [21] and provide an experimental evidence of a magnetic near-field channeling effect at microwave frequencies.

The geometry of the considered metamaterial lens is illustrated in Fig. 1(a). It consists of a periodic array of helical-shaped metallic wires infinitely extended along the $x$ and $y$ directions. The unit cell is a square (with spatial period $a = 10$ mm) and contains four helical-shaped wires (specifically, two right-handed helices and two left-handed helices) arranged in a checkerboard pattern [see Fig. 1(b)]. The radius of the helical-shaped wires is $R = 0.2a$, the helix pitch is $|p| = 0.45a$, the radius of the wires is $r_\mathrm{w} = 0.025a$, and the thickness of the lens along the $z$-direction is $L = 5a$. The free-space wavenumber and wavelength are respectively $k_0 = \omega/c$ and $\lambda_0 = 2\pi/k_0$.

Let us begin by discussing the transmission properties of the proposed metamaterial lens. To this end, we consider a scattering scenario where a TE plane wave propagating in the *xoz* plane ($k_y = 0$) with an electric field oriented along the *y*-direction illuminates the helical-shaped wire medium lens. It is worth noting that owing to the twofold rotational symmetry around the *z*-axis, the transmission properties are identical for a TE propagating in the *yoz* plane with the electric field oriented along the *x*-direction.

The transmission coefficient (*T*) of a metamaterial slab can be calculated using a nonlocal homogenization model that is reported in [21] and in the Supplementary Material. Figure 2(a) depicts the amplitude of *T* as a function of the transverse wave vector component $k_x$ for several frequencies of operation. The solid curves are associated with the nonlocal homogenization model, whereas the star symbols were obtained with the full-wave electromagnetic simulator CST Studio Suite [22]. One can see that there is a good agreement between the two sets of results for both amplitude (Fig. 2(a)) and phase (see the Supplementary Material) of *T*. Notably, Fig. 2(a)(*i*) predicts that for frequencies in the interval of 875 to 950 MHz, the transmission level exceeds unity ($|T| \geq 1$) over a significant range of the subwavelength (or evanescent) spatial spectrum ($1 < k_x c/\omega < 6$). This frequency interval is roughly centered around the Fabry-Perot resonance ($k_z^{\text{q-TEM}} L = \pi$, with $k_z^{\text{q-TEM}}$ the *z* component of the wavevector of the q-TEM mode) of the lens which occurs close to 900 MHz. On the other hand, even though the transmission level for the evanescent waves decreases substantially for frequencies below 875 MHz, the amplitude of *T* is still significantly different from 0 [see Fig. 2(a)(*ii*)].

The subwavelength channeling effect with enhanced transmission ($|T| > 1$) reported in Fig. 2(a) is rooted in the excitation of two bulk modes: a q-TEM propagating mode that travels across the lens (along the $z$-direction), as well as a guided mode, that is responsible for the resonant enhancement (i.e., the peaks of $|T|$), and that travels along the $x$-direction [see Fig. 2(b)] [21]. In fact, the resonant transmission enhancement is an important qualitative difference compared to the TM canalization regime characteristic of the standard wire medium lenses formed by straight parallel metallic wires. The standard wire medium lenses are operated in a regime where there is no excitation of guided modes travelling parallel to the interfaces (i.e., perpendicular to the wires), such that the transmission coefficient does not exhibit a resonant behavior and $|T| \approx 1$ for both propagating and evanescent spatial harmonics [23]. On the other hand, the resonant transmission enhancement provided by the helical-shaped wire medium lens may be useful in practical applications, mainly because it may enable to compensate for the exponential decay of the evanescent waves outside of the lens. However, it also has some drawbacks. In particular, for a finite-width lens, the reflection of the guided modes at the edges of the lens may corrupt the quality of the imaging. Despite the qualitative difference between the two channeling mechanisms, the results of Fig. 2(a) clearly indicate that somewhat analogous to the standard wire medium lens that enables channeling the electric near-field details of TM polarized waves, the helical-shaped wire medium lens may enable the transport of the subwavelength details of TE polarized waves.

To experimentally verify the subwavelength imaging capabilities of the proposed metamaterial lens, we fabricated a prototype of the helical-shaped wire medium that consists of a finite-size periodic array of $36 \times 36$ helical-shaped steel wires (i.e., $18 \times 18$ square unit cells) [see Fig. 3(a)]. The helical-shaped wires [see Fig. 3(b)] are embedded in a Styrofoam block whose relative permittivity is close to unity around the considered microwave frequencies. To assess the magnetic near-field channeling potential of the metamaterial lens, we used a printed antenna formed by an array of four small magnetic loops parallel to the interface plane as the near-field source [see Fig. 3(c)]. The loops are disposed in the form of a square and placed at a subwavelength distance of each other so that the resolution capabilities of the lens along both *x* and *y* directions can be fully evaluated. In addition, the loops are fed by currents in phase (the phase of the current does not play an important role here) and are placed at a subwavelength distance (about 2.5 mm) from the input interface of the metamaterial lens. The *z*-component of the magnetic field ($H_z$) was measured 2.5 mm above the output interface of the lens using a near-field scanning system based on a robotic arm with a round shielded loop probe connected to a vector network analyzer (R&S ZVB20) [see Fig. 3(d)].

The measured $H_z$ at the image plane and in the presence of the metamaterial lens is shown in Fig. 4(a) for different frequencies of operation in the range of $400-950$ MHz. In addition, the measured $H_z$ at the source plane and in the absence of the metamaterial lens is represented in Fig. 4(b) for $f = 897.63 \text{ MHz}$. It is clearly seen from Figs. 4(a)-(b) that, despite the subwavelength distance between the loops, the four magnetic loops are perfectly resolved at the image plane over a very broad frequency band (from 400 to 950 MHz). In contrast, the near-field radiation emitted by the loop array is completely imperceptible at the image plane when the metamaterial lens is removed [see Fig. 4(c)]. Thus, one can conclude that the imaging of the magnetic near field of the small loops is only possible because of the channeling effect provided by the helical-shaped wire medium lens [21].

In the Supplementary Material we also show the measured $H_z$ at the image plane and in the presence of the metamaterial lens for a different separation between the loops ($\Delta \approx 83 \text{ mm}$ rather than $\Delta \approx 41.5 \text{ mm}$). Such results are fully consistent with those of Fig. 4(a) and provide clear evidence that the channeling effect is not rooted in any special position of the near-field loops. It is worth noting that in both configurations ($\Delta \approx 41.5$ mm and $\Delta \approx 83$ mm) the loops are neither centered with the axis of the helices nor with the center of the unit cell.

A relevant aspect in the results shown in Fig. 4(a) is that the electromagnetic field distributions at the image plane are not fully symmetric along the *x* and *y* directions. In particular, the field intensity is not the same for all the four loops. The differences in the field intensities of the subwavelength loops (with perimeter in between $\lambda_0/10$ and $\lambda_0/25$) at the image plane are due to imperfections in the experimental setup. In particular, the distance between each loop and the lens input interface is not precisely the same. Due to the deeply subwavelength nature of the loops, their radiation properties (e.g., impedance) are highly sensitive to small disturbances. As a result, even minimal distance discrepancies can lead to significant differences in the field amplitude that reaches the input interface of the lens. These asymmetries are accurately reproduced by the lens at the image plane.

In order to confirm the experimental results of Fig. 4, the electromagnetic response of the same setup was simulated using the full-wave simulator [22]. Figure 5(a)-(b) depicts the simulated *z*-component of the magnetic field, $H_z$, at the image plane and in the presence of the metamaterial lens (Fig. 5(a)), and at the source plane without the metamaterials lens (Fig. 5(b)). The simulated fields are consistent with the experimental results of Fig. 4(a-b), and in particular it can be seen from Fig. 5(a) that the four loops are clearly resolved at the image plane. Figure 5(c) provides a clear evidence of the magnetic near-field channeling effect performed by the helical-shaped wire metamaterial lens. As one can see, the radiation of the loops is clearly channeled by the metamaterial lens along the axial direction (*z*-direction) with negligible lateral spreading.

Notably, the magnetic field amplitude at the image plane in the presence of the metamaterial lens is different from zero on spatial regions significantly far from the position of the four loops. This property can be seen in the experimental results (Fig. 4(a)) for certain frequencies (e.g., $f = 897.63 \text{ MHz}$, $f = 911.96 \text{ MHz}$, and $f = 950 \text{ MHz}$), as well in the simulation results (Fig. 5(a)). This happens because of the excitation of the guided modes travelling along the $x$ and $y$ directions of the metamaterial lens and due to the diffraction effects at the edges of the lens. In principle, microwave absorbers could be placed on the sides of the metamaterial lens to avoid these field interferences.

Due to the limited resolution of the metamaterial lens, the helical-shaped wire medium lens is unable to channel the finest details of the small magnetic loops to the image plane [see Fig. 4(a) and Fig. 5(a)]. For example, the magnetic field distribution of each loop is formed by a central maximum (ring) with another maximum with opposite phase surrounding it (annulus) [see Fig. 4(b)], but this structure is not perfectly reproduced at the image plane [see Fig. 4(a) and Fig. 5(a)]. The resolution of the helical-shaped wire medium lens is about twice the lattice period [21] (which corresponds to a resolution between $\lambda_0/40$ and $\lambda_0/16$ in the considered frequency range), whereas the distance between the two maxima of the magnetic field for each of the loops is much smaller than $2a$. For completeness, it is worth mentioning that an alternative version of the lens with two coaxial, interlaced, and non-connected helices with opposite handedness per unit cell (i.e., with half spatial period) may potentially improve resolution by a factor of two.

In conclusion, we have experimentally verified that a magnetic wire medium lens based on a racemic array of helical-shaped metallic wires enables the channeling of the subwavelength details of the normal component of the magnetic field of near-field sources over a broad bandwidth. It was experimentally demonstrated that the magnetic wire medium lens effectively behaves as the magnetic analogue of the conventional wire medium lens formed by parallel straight metallic wires, enabling the sampling of the magnetic near-field distribution at the input interface of the lens and its transport to the output interface with little distortion. In principle, the design of this magnetic wire medium lens can be scaled to terahertz and infrared frequencies [9]. The reported magnetic channeling effect with subwavelength resolution provided by the proposed metamaterial lens may have important impact on several areas, such as in magnetic resonance imaging (MRI) [5, 24] acting as a magnetic flux-guiding medium with super-resolution, in near-field wireless power transfer (NF-WPT) [25-26] by promoting a strong evanescent wave coupling and enhancing the power transfer efficiency, and in sensing of magnetic objects separated by subwavelength distances [27].

**Supplemental Material:**

Supplemental Material with the (*i*) nonlocal homogenization model that characterizes the electromagnetic response of the proposed helical-shaped wire metamaterial; (*ii*) the plane-wave scattering problem based on the nonlocal homogenization approach; (*iii*) further experimental results of the measured *z*-component of the magnetic field.

**Acknowledgments:**

This work was funded by Instituto de Telecomunicações (IT) under project HelicalMETA - UIDB/50008/2020. T. A. Morgado acknowledges FCT for research financial support with reference CEECIND/04530/2017 under the CEEC Individual 2017, and IT-Coimbra for the contract as an assistant researcher with reference CT/No. 004/2019-F00069. G. L. João acknowledges support by IT-Coimbra under a scientific initiation grant within the scope of the project HelicalMETA - UIDB/50008/2020. R. A. M. Pereira acknowledges support by FCT, POCH, and the co-financing of Fundo Social Europeu under the fellowship SFRH/BD/145024/2019. D. E. Fernandes acknowledges support by FCT, POCH, and the co-financing of Fundo Social Europeu under the fellowship SFRH/BPD/116525/2016 and by IT-Lisbon under the research contract with reference C-0042-22. S. Lannebère acknowledges financial support by IT-Coimbra under the research contract with reference DL 57/2016/CP1353/CT0001.

# Figures

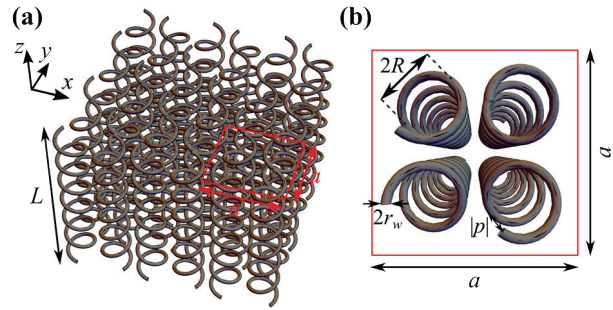

**Figure 1.** (a) Geometry of the helical-shaped wire medium slab with thickness $L$: a racemic array of helical-shaped metallic wires periodic along the $x$ and $y$ directions (with lattice period $a$ along both directions). (b) Unit cell of the metamaterial which includes two right-handed helices and two left-handed helices arranged in a checkerboard pattern.

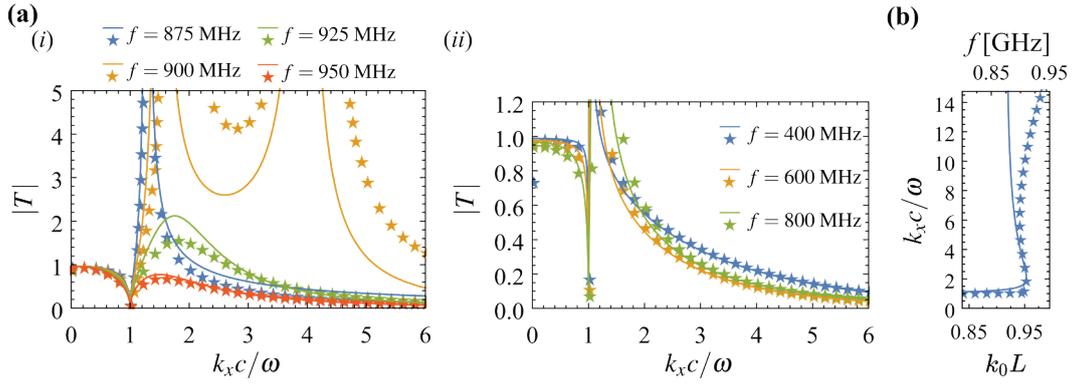

**Figure 2.** (a) Amplitude of the transmission coefficient $T$ as a function of the normalized transverse component of the wave vector $k_x$ for different frequencies of operation. (b) Normalized propagation constant $k_x c/\omega$ of the low-frequency guided mode supported by the helical-shaped wire metamaterial as a function of the normalized frequency $k_0 L$. Solid lines: nonlocal homogenization results; star symbols: full-wave simulations [22]. The geometrical parameters in (a) and (b) are, $a = 10$ mm, $R = 0.2a$, $r_w = 0.025a$, $|p| = 0.45a$, and $L = 5a$.

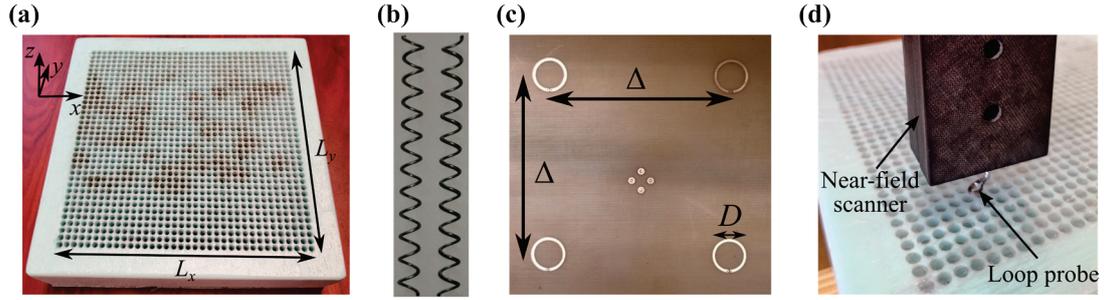

**Figure 3.** (a) Photo of the prototype of the magnetic uniaxial wire medium lens formed by $36 \times 36$ helical-shaped steel wires with $R = 2$ mm, $|p| = 4.5$ mm, and $r_w = 0.25$ mm. The metamaterial lens has dimensions $L_x = L_y = 18a = 180$ mm and thickness along the $z$-direction $L = 5a = 50$ mm. (b) Photo of the two different metamaterial inclusions: a right-handed and a left-handed helical-shaped wire. (c) Photo of the excitation near-field antenna array formed by four small magnetic loops with $D \approx 9.5$ mm ($\lambda_0/80 \leq D \leq \lambda_0/30$ in the considered $[400-950]$ MHz frequency band) and separated by a subwavelength distance $\Delta \approx 41.5$ mm ($\lambda_0/18 \leq \Delta \leq \lambda_0/8$ in the considered frequency range). (d) Photo of the experimental setup with the metamaterial lens and the near-field scanning system with a round shielded loop probe.

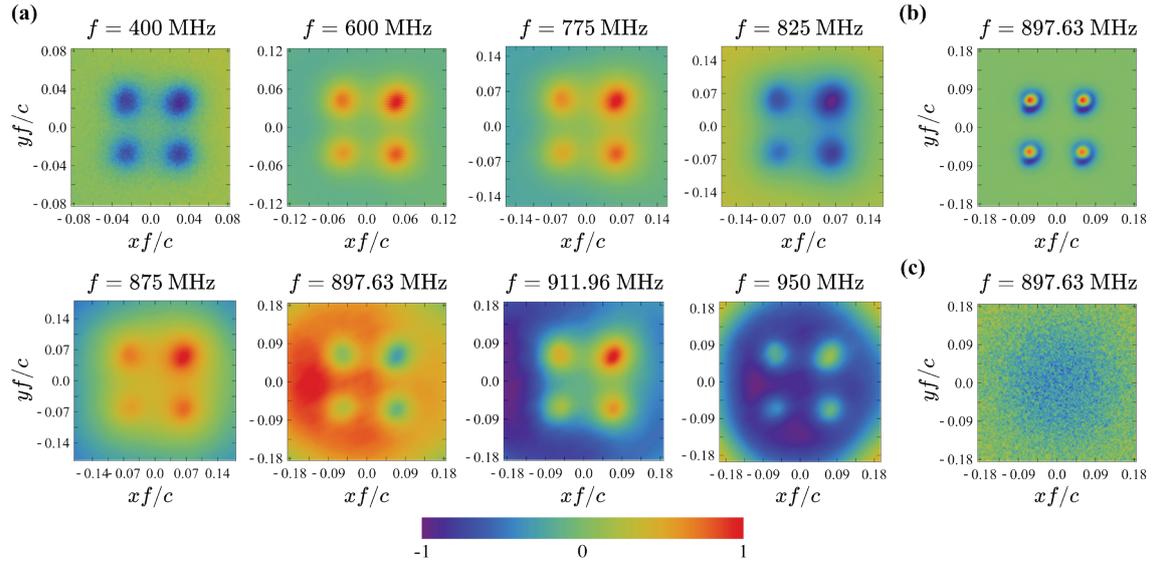

**Figure 4.** Experimental time snapshots of the normalized *z*-component of the magnetic field for different frequencies; (a) at the image plane (about 2.5 mm above the output interface of the lens); (b) at the source plane (about 2.5 mm above the loop-antenna array) and in the absence of the metamaterial lens; (c) at the image plane as in (a) but without the metamaterial lens.

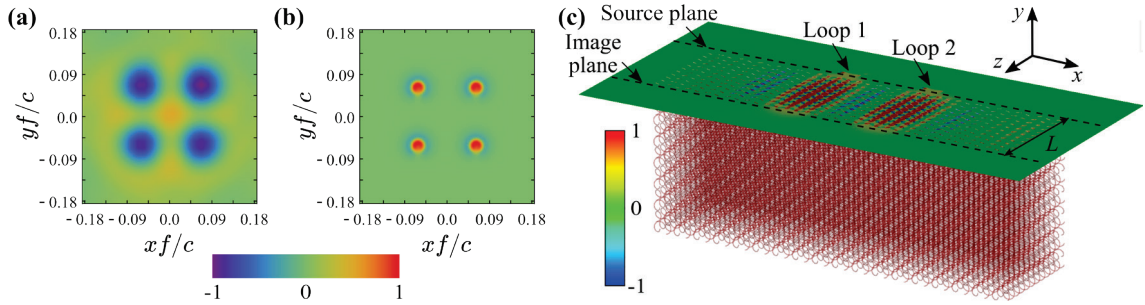

**Figure 5.** Time snapshots of the normalized $z$-component of the magnetic field for $f = 897.63$ MHz, obtained using the full-wave electromagnetic simulator CST Studio Suite [22]; (a) at the image plane (about 5 mm from the output interface of the lens); (b) at the source plane (about 5 mm from the loop-antenna array) without the metamaterial lens. (c) At a $xoz$ plane that cuts the metamaterial lens close to the middle of the lens and passes through the center of two loops. The remaining parameters of (a), (b) and (c) are the same as in Fig. 4.

# Supplemental Material for the Manuscript

# "Experimental verification of the magnetic near-field channeling using a helical-shaped wire medium lens"

Tiago A. Morgado[1*], Guilherme L. João[1], Ricardo A. M. Pereira[2], David E. Fernandes[1] and Sylvain Lannebère[1]

[1]*Instituto de Telecomunicações and Department of Electrical Engineering, University of Coimbra, 3030-290 Coimbra, Portugal*

[2]*Instituto de Telecomunicações and Department of Electronics, Telecommunications and Informatics, University of Aveiro, 3810-193 Aveiro, Portugal*

E-mail: tiago.morgado@co.it.pt

In the supplementary note *A*) we present the nonlocal homogenization model that characterizes the electromagnetic response of the racemic array of helical-shaped metallic wires. In the supplementary note *B*), we solve the plane wave scattering problem using a combination of mode matching and additional boundary conditions. Finally, in the supplementary note *C*) we provide further experimental results of the measured *z*-component of the magnetic field.

## *A. Nonlocal homogenization model*

The electromagnetic response of the considered metamaterial can be accurately characterized using effective medium theory [1-2]. It is demonstrated in [1-2] that the racemic helical-shaped wire medium behaves as a spatially dispersive (nonlocal) uniaxial non-bianisotropic electric and magnetic material described by diagonal effective permittivity and permeability tensors. In particular, the relative effective permeability tensor [1-2] is given by

$$\overline{\overline{\mu}}_r = \hat{\mathbf{u}}_x\hat{\mathbf{u}}_x + \hat{\mathbf{u}}_y\hat{\mathbf{u}}_y + \mu_{zz}\hat{\mathbf{u}}_z\hat{\mathbf{u}}_z, \quad \mu_{zz} = \left(1 + \frac{A^2 k_0^2}{\frac{k_0^2}{\beta_{p1}^2} - \frac{k_z^2}{\alpha\beta_{p2}^2}}\right)^{-1}, \tag{S1}$$

where $k_0 = \omega/c$, with $\omega$ being the operation angular frequency and $c$ the light speed in free space, $A = \pi R^2/|p|$, and $\beta_{p1}$ and $\beta_{p2}$ are wave number parameters that only depend on the geometry of the material and are given by:

$$\beta_{p1} = 4\pi\sqrt{\frac{p^2}{C_0 p^2 V_{\text{cell}} + 8C_1 \pi^2 R^2 V_{\text{cell}}}}, \tag{S2}$$

$$\beta_{p2} = 4\pi\sqrt{\frac{1}{C_0 V_{\text{cell}}}}, \tag{S3}$$

where $V_{\text{cell}} = a^2|p|$ is the volume of the unit cell, and $C_0$ and $C_1$ are geometrical parameters [3]. The parameter $\alpha$ is an adjustable coefficient used to obtain a better matching with the full wave simulations. It depends on the geometry of the metamaterial [2]. For the geometry of the main text with $R = 0.2a$, $r_w = 0.025a$, and $|p| = 0.45a$, we have $C_0 a = 6.094$, $C_1 a = 4.474$, $\beta_{p1} a \simeq 2.15$, $\beta_{p2} a \simeq 7.59$, and $\alpha = 0.695$.

---

[*] To whom correspondence should be addressed: E-mail: tiago.morgado@co.it.pt

The dispersion characteristic of the TE-$z$ plane-wave like modes (with electric field parallel to the $xoy$ plane) supported by the helical-shaped wire medium is given by [1-2]

$$k_0^2 \varepsilon_{\text{t}} - (1/\mu_{zz})k_{\text{t}}^2 - k_z^2 = 0, \tag{S4}$$

where $k_{\text{t}}^2 = k_x^2 + k_y^2$ and $\varepsilon_{\text{t}} = \varepsilon_{xx} = \varepsilon_{yy} = 1 + (2\pi R)^2/(V_{\text{cell}} C_1)$. The characteristic equation (S4) reduces to a polynomial of degree 2 in the variables $k_0^2$ and $k_z^2$. Thus, for each wave vector there are two different TE-$z$ plane-wave modes propagating in the helical-shaped wire medium. The emergence of an extra mode is related to the nonlocal response of the helical-shaped wire medium, which is implicit from the dependence of the effective permeability [Eq. (SS1)] on the wave vector.

## B. Scattering problem

The scattering properties of a helical-shaped wire metamaterial slab can be studied using the nonlocal homogenization model [Eq. (S1)]. The metamaterial is assumed infinite and periodic along the *x* and *y* directions, and finite, with thickness *L,* along the *z* direction. The incident plane wave propagates in the *xoz* plane ($k_y = 0$), with the electric field polarized along the *y*-direction, parallel to the interface [see Fig. S1]. Thus, the electric field in all the regions can be written as (the *x*-variation of the fields is suppressed):

$$E_y^{(1)} = E_y^{inc}(e^{\gamma_0 z} + \rho e^{-\gamma_0 z}), \quad z > 0$$
$$E_y^{(2)} = A_1^+ e^{-jk_z^{\text{q-TEM}} z} + A_1^- e^{+jk_z^{\text{q-TEM}} z} + A_2^+ e^{-jk_z^{\text{TE}} z} + A_2^- e^{+jk_z^{\text{TE}} z}, \quad -L < z < 0, \quad \text{(S5)}$$
$$E_y^{(3)} = E_y^{inc} T e^{\gamma_0 (z+L)}, \quad z < -L$$

where, $E_y^{inc}$ is the incident electric field, $\gamma_0 = \sqrt{k_x^2 - \omega^2 \varepsilon_0 \mu_0}$ is the free-space propagation constant, $k_x = \omega\sqrt{\varepsilon_0 \mu_0} \sin\theta_i$, and $\rho$ and $T$ are the reflection and transmission coefficients, respectively. The propagation constants $k_z^{\text{q-TEM,TE}}$ [calculated by solving Eq. (S4) with respect to $k_z$] and the amplitudes $A_{1,2}^{\pm}$ are associated with the two different modes excited inside the metamaterial slab. The plus and minus superscripts correspond to the two counterpropagating waves excited in the slab. For each plane wave with electric field of the form $E = E_0 e^{-j\mathbf{k}\cdot\mathbf{r}} \hat{\mathbf{u}}_y$, the corresponding magnetic field is given by

$$\mathbf{H} = \frac{E_0}{\eta_0 k_0}\left(-k_z \hat{\mathbf{u}}_x + \frac{k_x}{\mu_{zz}} \hat{\mathbf{u}}_z\right) e^{-j\mathbf{k}\cdot\mathbf{r}}. \quad \text{(S6)}$$

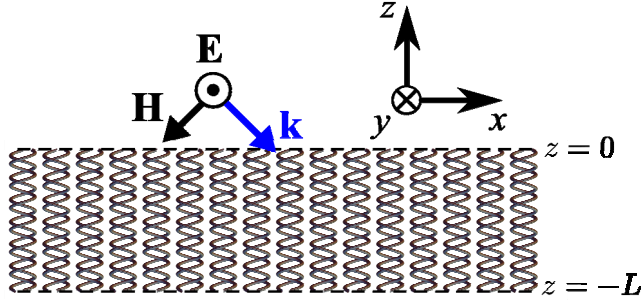

**Figure S1.** Geometry of a helical-shaped wire metamaterial slab (infinite and periodic along the *x* and *y* directions) with thickness *L* along *z*. The plane of incidence is the *xoz* plane and the wave is TE-*z* polarized ($\mathbf{k}^{inc} = (k_x, 0, k_z^{inc})$, $\mathbf{E}^{inc} = E^{inc}\hat{\mathbf{u}}_y$).

The reflection and transmission coefficients are determined by the following set of boundary conditions [2]:

$$E_y \text{ and } H_x \text{ are continuous at } z = 0 \text{ and } z = -L, \quad \text{(S7a)}$$

$$H_z \text{ is continuous at } z = 0 \text{ and } z = -L \quad \text{(S7b)}$$

The first set of boundary conditions (Eq. (S7a)) are the conventional ones which imply the continuity of the tangential components of the electromagnetic fields at the interfaces, as there are no surface currents nor surface magnetization. The second set of boundary conditions [Eq. (S7b)] is necessary to model the effects of spatial dispersion and the behavior of the microscopic currents flowing along the helical-shaped metallic wires. It establishes that the normal component of the magnetic field ($H_z$) is continuous at both interfaces [2]. Such an additional boundary condition (ABC) ensures that the macroscopic magnetic current density ($\mathbf{J}_m = j\omega(\mathbf{B} - \mu_0 \mathbf{H})$) along the *z*-direction vanishes at the interfaces [2]. By imposing the two sets of boundary conditions, we obtain a linear system of equations which can be solved numerically with respect to the unknowns, and which thereby determine the transmission and reflection coefficients. In particular, one can obtain the amplitude (Fig. 2(a) of the main text) and phase (Fig. S2) of the transmission coefficient.

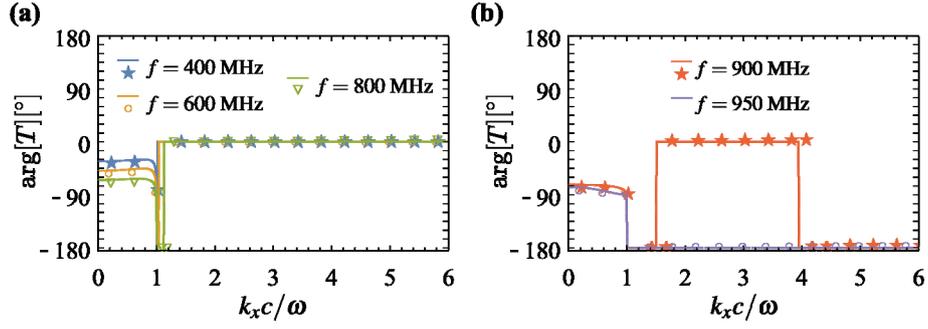

**Figure S2.** Phase of the transmission coefficient $T$ as a function of the normalized transverse component of the wave vector $k_x$ for different frequencies of operation. Solid lines: nonlocal homogenization results; star symbols: full-wave simulations [4]. The geometrical parameters in (a) and (b) are, $a = 10$ mm, $R = 0.2a$, $r_w = 0.025a$, $|p| = 0.45a$, and $L = 5a$.

Figure S2 depicts the phase of the transmission coefficient $T$ as a function of $k_x$ in the frequency range $[400-950]$ MHz. The solid lines correspond to the nonlocal homogenization results and the discrete symbols to full-wave simulations [4]. It is clearly seen from Fig. S2 that the results obtained from the two different approaches concur well. As one can see, for frequencies between $400-800$ MHz the phase of $T$ is equal to $0°$ for the evanescent (or near-field) spatial spectrum ($1 < k_x c/\omega < 6$). For intermediate frequencies close to the Fabry-Perot resonance ($900$ MHz) the phase of $T$ for the evanescent spectrum alternates between $0°$ and $-180°$. On the other hand, for frequencies above the Fabry-Perot resonance the phase of $T$ is equal to $-180°$ for the evanescent spatial harmonics ($k_x c/\omega > 1$).

## C. Additional experimental results

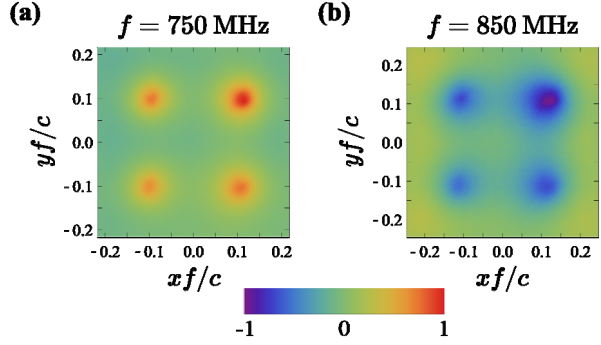

**Figure S3.** Experimental time snapshots of the normalized *z*-component of the magnetic field at the image plane (about 2.5 mm above the output interface of the lens) when the four small magnetic near-field loops are separated by a subwavelength distance $\Delta \approx 83$ mm, i.e. (a) $\Delta \approx 0.2\lambda_0$; (b) $\Delta \approx 0.235\lambda_0$.

Figure S3 shows additional experimental results of the measured $H_z$ at the image plane in the presence of the metamaterial lens for a different separation between the loops ($\Delta \approx 83$ mm) from that considered in the main text ($\Delta \approx 41.5$ mm). Consistent with the results of Fig. 4(a) of the main text, Fig. S3 further demonstrates that the helical-shaped wire medium lens enables channeling the magnetic near-field of the four small loops. Moreover, the results of Fig. S3 demonstrate that the reported near-field channeling effect does not just happen for a specific configuration, namely for a specific position of the loops relatively to the helical-shaped wires of the lens. It is important to point out here that the loops are neither centered with the axial axis of the helices nor with the center of the unit cell, both for the configuration with $\Delta \approx 41.5$ mm (Fig. 4(a) of the main text) and for the configuration with $\Delta \approx 83$ mm (Fig. S3).